\documentclass[11pt,twoside]{article}
\usepackage{asp2010}

\resetcounters

\begin{document}

\title{Asteroseismic analysis of the CoRoT target HD~169392}
\author{Savita Mathur$^{*, 1,2}$, H. Bruntt$^{3}$, C. Catala$^{4}$, O. Benomar$^{5}$, G. R. Davies$^{2}$, R. A. Garc\'ia$^{2}$, D. Salabert$^{6}$, J. Ballot$^{7,8}$, B. Mosser$^{4}$, C. R\'egulo$^{9,10}$, W. J. Chaplin$^{11}$, Y. Elsworth$^{11}$, R. Handberg$^{11, 3}$, S. Hekker$^{12,11}$, L. Mantegazza$^{13}$, E. Michel$^{4}$, E. Poretti$^{13}$, M. Rainer$^{13}$, I. W. Roxburgh$^{14}$, R. Samadi$^{4}$, M. Steslicki$^{15,1}$, K. Uytterhoeven$^{9,10}$, G. A. Verner$^{11}$, M. Auvergne$^{4}$, A. Baglin$^{4}$, S. Barcel\'o Forteza$^{9,10}$, F. Baudin$^{16}$, and T. Roca Cort\'es$^{9,10}$\\
{$^*$} Affiliations are given at the end of the paper}

\begin{abstract}
The satellite CoRoT (Convection, Rotation, and planetary Transits) has provided high-quality data for almost six years. We show  here the asteroseismic analysis and modeling of HD169392A, which belongs to a binary system weakly gravitationally bound as the distance between the two components is of $\sim$4250 AU. The main component, HD169392A, is a G0IV star with a magnitude of 7.50 while the second component is a G0V-G2IV star with a magnitude of 8.98. This analysis focuses on the main component, as the secondary one is too faint to measure any seismic parameters. A complete modeling has been possible thanks to the complementary spectroscopic observations from HARPS, providing $T_{\rm eff}$=5985$\pm$60K, $\log g$=3.96$\pm$0.07, and [Fe/H]=-0.04$\pm$0.10. 
\end{abstract}

\section{Introduction}

Stellar physics are undergoing a tremendous revolution thanks to the large amount of data obtained using photometry, spectroscopy, and interferometry. One major breakthrough the last few years concerned asteroseismology thanks to the exquisite data provided by space-based missions such as CoRoT \citep[Convection, Rotation, and planetary Transits][]{2006cosp...36.3749B} and {\it Kepler} \citep{2010Sci...327..977B}. With asteroseismology we can probe the internal structure of the stars \citep{2011Natur.471..608B,2012ApJ...749..152M} as well as their dynamics \citep{2012Natur.481...55B,2012ApJ...756...19D}. These missions also showed how the search for exo-planet transits and asteroseismology are tightly linked \citep[e.g.][]{2012ApJ...746..123H}. It is undeniable that radius, mass, and age are obtained more accurately with seismic analysis than with classical methods \citep[e.g.][]{2013ApJ...763...49D,2012ApJ...748L..10M}, thus providing a better estimate of the radii of the planet and the age of planetary systems in the case of planet host stars. 

\noindent We present here the analysis a solar-like star, HD~169392A observed by CoRoT during the third long run in the central direction of the galaxy (LRc03). We obtained continuous observations for 91.2 days. The time series utilised in this work were prepared by the CoRoT Data Center (CDC) \citep{2007astro.ph..3354S}. By combining the asteroseismic analysis with spectroscopic observations, we derive the stellar parameters of the stars.

\section{Asteroseismic analysis}

The data were obtained with a regular cadence of 32 s in the heliocentric frame. The time series is shown in Figure 1 of \citet{2013A&A...549A..12M} where we can see two jumps at 40.3 and 80.05 days. These jumps are probably of instrumental origin.

\noindent  We applied the ÒinpaintingÓ algorithm \citep[e.g.][]{2010arXiv1003.5178S} to replace the perturbed data due to the South Atlantic Anomaly and remove the peaks corresponding to the daily harmonics. Three components of the background were fitted, including different scales of granulation, with a maximum likelihood estimator \citep[e.g.][]{2011ApJ...741..119M}.

\subsection{Acoustic-mode fitting}

Nine teams estimated the mode parameters of HD~169392A using  Maximum Likelihood Estimation (MLE), maximum a priori (MAP), and Markov Chain Monte Carlo (MCMC). Minimal and maximal lists of frequencies were built based on the Peirce criterion \citep{1852AJ......2..161P}. The results of the fitting analysis can be found in \citet{2013A&A...549A..12M}. The l=3 modes were fitted and belong to the maximal list.
We notice the presence of an avoided crossing suggesting the presence of a mixed mode \citep[e.g.][]{1975PASJ...27..237O}.
By applying the formalism developed by \citet{2012A&A...540A.143M}, we find a gravity period spacing $\Delta \Pi$=476.9$\pm$4.3s. We report 2 mixed modes at 816 and 1336 $\mu$Hz. The first one is surrounded by the orbital harmonics of CoRoT while the second one has a very weak signal so it was not fitted.

\noindent No signature of surface rotation was found in the light curve. We performed multiple fits  for a fixed rotational splitting and a range of fixed angles. Two maxima of almost equally high probability were obtained in the joint-posterior probability density function. So two different values for the joint-parameters are possible: an inclination angle between 20 and 40$^\circ$ with a rotational splittings of 0.4-1.0$\mu$Hz or between 55 and 86$^\circ$ for a rotational splittings of 0.2-0.5$\mu$Hz.

\subsection{Modeling}

Using the spectroscopic constraints and the asteroseismic parameters, we modeled HD169392A with the Asteroseismic Modeling Portal \citep[AMP,][]{2009ApJ...699..373M}, the infrared flux method \citep[IRFM,][]{2010A&A...512A..54C}, and the IRFM combined with grid-based method \citep{2012ApJ...757...99S}. The results agree within uncertainties. For the AMP model, we computed the frequency ratios, r$_{01}$ and r$_{10}$ as in \citet{2003A&A...411..215R}. They are sensitive to the core structure of the star. The comparison of the frequency ratios from the observations and the models (Figure 1) shows a good agreement ($\chi^2$ of ~8 and 4) suggesting that the models reproduce well the surface and the interior of the star.

\begin{figure}[htbp]
\begin{center}
\includegraphics[angle=90, width=10cm]{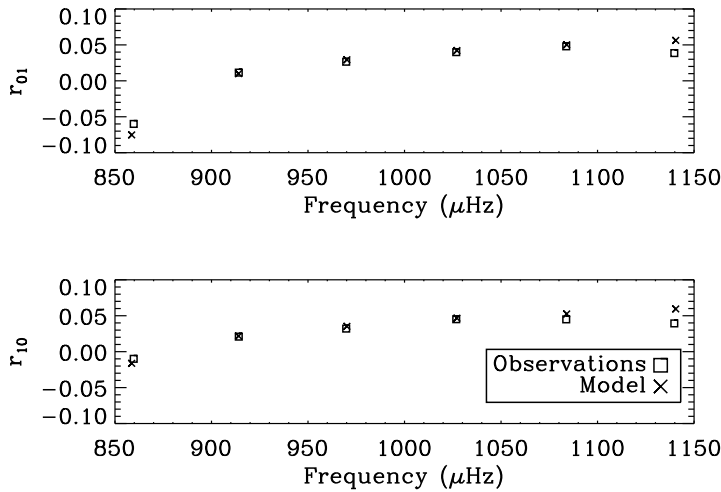}
\caption{Frequency ratios of HD169392A as defined in \citet{2003A&A...411..215R} in the range 850 to 1150$\mu$Hz. The square symbols represent the observations while the crosses represent the best-fit model from AMP.}
\label{default}
\end{center}
\end{figure}

\section{Conclusions}
We extracted the mode parameters for l=0, 1, 2, and 3 for HD169392A. The signal-to-noise ratio of the second component was too weak to study the oscillations.
The analysis of splittings and inclination angle gives two possible solutions: one with splitting and inclination angle of 0.4-1.0$\mu$Hz and 20-40$^\circ$, the other with 0.2-0.5$\mu$Hz and 55-86$^\circ$ .
The Asteroseismic Modeling Portal (AMP) gives a mass of 1.15$\pm$0.01M$_\odot$, a radius of 1.88$\pm$0.02R$_{\odot}$, and an age of 4.33$\pm$0.12Gyr, for a given physics \citep{2012ApJ...748L..10M}. The error bars are the internal uncertainties. These results agree with results obtained from other modeling methods. We conclude that this star is a subgiant that has exhausted its central hydrogen and has no convection core.
The study of the frequency ratios, r$_{01}$ and r$_{10}$, shows good agreement between the best-fit model from AMP and the observations, which implies that the properties of the core are well represented by the model.

\acknowledgements The CoRoT space mission has been developed and is operated by CNES, with contributions from Austria, Belgium, Brazil, ESA (RSSD and Science Program), Germany and Spain. RAG acknowledges the support given by the French PNPS program. RAG and SM acknowledge the CNES for the support of the CoRoT activities at the SAp, CEA/Saclay and the support of the European Community's Seventh Framework Programme (FP7/2007-2013) under grant agreement no. 269194 (IRSES/\-ASK). DS acknowledges the support from CNES. SH acknowledges financial support from the Netherlands Organization for Scientific Research (NWO).  LM, EP, and MR acknowledge financial support from the PRIN-INAF 2010 (Asteroseismology: looking inside the stars with space- and ground-based observations). IWR acknowledges support from the Leverhulme. Foundation under grant 2012-035/4. KU acknowledges financial support by the Spanish National Plan of R\&D for 2010, project AYA2010- 17803. CR, SBF and TRC wish to thank financial support from the Spanish Ministry of Science and Innovation (MICINN) under the grant AYA2010-20982- C02-02. NCAR is partially funded by the National Science Foundation. SM acknowledges the support of the University of Tokyo. 

Affiliations: 
{$^1$High Altitude Observatory, NCAR, P.O. Box 3000, Boulder, CO 80307, USA.}
{$^{2}$La\-bo\-ra\-toi\-re AIM, CEA/DSM -- CNRS - Universit\'e Paris Diderot -- IRFU/SAp, 91191 Gif-sur-Yvette Cedex, France.}
{$^3$Danish AsteroSeismology Centre, Department of Physics and Astronomy, University of Aarhus, 8000 Aarhus C, Denmark}
{$^{4}$LESIA, UMR8109, Universit\'e Pierre et Marie Curie, Universit\'e Denis Diderot, Obs. de Paris, 92195 Meudon Cedex, France.}
{$^{5}$Sydney Institute for Astronomy, School of Physics, University of Sydney, NSW 2006, Australia.}
{$^6$Laboratoire Lagrange, UMR7293, UniversitŽ de Nice Sophia-Antipolis, CNRS, Observatoire de la C\^ote dÕAzur, 06304 Nice Cedex4, France}
{$^7$Institut de Recherche en Astrophysique et Plan\'etologie, Universit\'e de Toulouse, CNRS, 14 avenue E. Belin, 31400 Toulouse, France.}
{$^8$Universit\'e de Toulouse, UPS-OMP, IRAP, 31400 Toulouse, France.}
{$^{9}$Universidad de La Laguna, Dpto de Astrof\'isica, 38206, Tenerife, Spain.}
{$^{10}$Instituto de Astrof\'\i sica de Canarias, 38205, La Laguna, Tenerife, Spain.}
{$^{11}$School of Physics and Astronomy, University of Birmingham, Edgbaston, Birmingham B15 2TT, UK.}
{$^{12}$Astronomical Institute "Anton Pannekoek", University of Amsterdam, PO Box 94249, 1090 GE Amsterdam, The Netherlands.}
{$^{13}$INAF - Osservatorio Astronomico di Brera, via E. Bianchi 46, 23807, Merate (LC), Italy}
{$^{14}$Astronomy Unit, Queen Mary University of London, Mile End Road, London E1 4NS, UK}
{$^{15}$Space Research Center, Polish Academy of Sciences, 51-622, Kopernika 11, Wroc?aw, Poland}
{$^{16}$Institut dÕAstrophysique Spatiale, UMR8617, Universit\'e Paris XI, Batiment 121, 91405 Orsay Cedex, France}

\bibliographystyle{asp2010} 
\bibliography{/Users/Savita/Documents/BIBLIO_sav.bib}

\end{document}